\documentclass[pra,twocolumn,floatfix,showpacs]{revtex4}

\usepackage{graphicx}

\begin{document}
\date{\today}
\author{S. Trotzky$^{1,\dagger}$}
\author{L. Pollet$^{2,3,\dagger}$}
\author{F. Gerbier$^4$}
\author{U. Schnorrberger$^1$}
\author{I. Bloch$^{1,5}$}
\author{N.V. Prokof'ev$^{2,6}$}
\author{B. Svistunov$^{2,6}$}
\author{M. Troyer$^3$}

\affiliation{
  $^1$ Institut f\"ur Physik, Johannes Gutenberg-Universit\"at, 55099 Mainz, Germany\\
  $^2$ Department of Physics, University of Massachusetts, Amherst, MA 01003, USA\\
  $^3$ Theoretische Physik, ETH Zurich, CH-8093 Zurich, Switzerland\\
  $^4$ Laboratoire Kastler Brossel, ENS, UPMC, CNRS, 24 rue Lhomond, 75005 Paris, France\\
  $^5$ Max-Planck-Institut f\"ur Quantenoptik, 85748 Garching, Germany\\
  $^6$ Russian Research Center ``Kurchatov Institute'', 123182 Moscow, Russia	\\
  $^\dagger$ These authors contributed equally to this work
}

\title{Suppression of the critical temperature for superfluidity near the Mott transition: validating a quantum simulator}

\begin{abstract}
Ultracold atomic gases in optical lattices have proven to be a controllable, tunable and clean implementation of strongly interacting quantum many-body systems. An essential prospect for such quantum simulators is their ability to map out the phase diagram of fundamental many-body model Hamiltonians. However, the results need to be validated first for representative benchmark problems via state-of-the-art numerical methods of quantum many-body theory. Here we present the first {\em ab-initio} comparison between experiments and quantum Monte Carlo simulations for strongly interacting Bose gases on a lattice for large systems (up to $N \simeq 3\times10^5$ particles). The comparison has enabled us to perform thermometry for the interacting quantum gas and to experimentally determine the finite temperature phase diagram for bosonic superfluids in an optical lattice. Our results reveal a downshift of the critical temperature as the transition to the Mott insulator is approached.
\end{abstract}

\pacs{03.75.Hh, 03.75.Lm, 37.10.Jk, 75.40.Mg}

\maketitle

Ultracold bosonic atoms in optical lattices have sparked investigations of strongly correlated many-body quantum phases with ultracold atoms~\cite{jaksch2005,lewenstein2007,bloch2008} that are now at the forefront of current research~\cite{greiner2002,paredes2004,kinoshita2004,hadzibabic2006}. For increasing interactions between the particles, a bosonic superfluid (SF) converts into a Mott insulator (MI), with dramatically different properties~\cite{fisher1989,jaksch1998,sachdevbook}. So far, several of the characteristic ground state properties of the systems either in the SF or in the MI have been measured~\cite{greiner2002,stoferle2004,gerbier2005a,gerbier2006a,folling2006,campbell2006,spielman2007,mun2007,spielman2008,guarrera2008}. However, up to now, no finite temperature phase diagram of the system~\cite{sheshadri1993a,elstner1999a,dickerscheid2003a,demarco2005a,pupillo2006a,blakie2007a,capogrosso2007} could be determined experimentally. Furthermore, a series of papers has questioned the analysis of the momentum distributions observed in the experiments~\cite{gerbier2005a}, arguing that the temperature could be higher than anticipated~\cite{diener2007,kato2008}. 

In this paper, we present for the first time a direct comparison of an experiment with ultracold bosons in an optical lattice with ab-initio finite temperature, quantum Monte Carlo (QMC) simulations. The simulations are performed for realistic trapping potentials and particle numbers without free parameters and include important effects such as the finite time-of-flight (TOF) and finite imaging resolution~\cite{gerbier2008}. We rely on the sudden appearance of narrow interference peaks (or a sudden change in the peak width) on top of a broader thermal background to detect the onset of long-range phase coherence~\cite{kollath2004}, equivalent to the onset of superfluidity in the three-dimensional system. Recently, the question of whether a purely normal cloud could also give rise to such narrow peaks has been raised theoretically, thus questioning the interpretation of this interference pattern as a signature for the appearance of a superfluid component~\cite{kashurnikov2002,kato2008}. So far, this question could not be settled due to a lack of an accurate thermometry method in the optical lattice potential~\cite{muradyan2008,mckay2009a}. Here, the direct comparison of the experimental data and the simulations allows to determine a temperature for the ultracold lattice gas. Theoretically, this temperature can also be computed from the measured initial temperature, assuming adiabaticity during the loading of the lattice~\cite{rey2006a,ho2007a,gerbier2007,pollet2008a}. We find good agreement between the two approaches, and conclude that the system is prepared almost adiabatically, up to small non-adiabatic heating effects of technical origin which we quantify. For weak interactions, we observe the sudden appearance of sharp interference peaks near a critical temperature $T_{c}$, confirming the sudden appearance of a condensate as expected for weakly interacting gases. For stronger interactions, yet below the Mott transition, although the appearance of a phase-coherent component with decreasing temperature is still clear, a smoother evolution is observed, with broad interference peaks also present in the normal fluid (NF) phase~\cite{kato2008,zhou2009}. Finally, we establish a method to extract the critical temperature $T_c$ for superfluidity in the lattice in both regimes. As we approach the Mott transition, a downshift of $T_c$ that cannot be accounted for by single-particle effects is observed.

\emph{Bose-Hubbard Model with External Confinement.}
We consider a system of ultra-cold bosons in a three-dimensional lattice potential with an additional harmonic trap. At low enough temperatures, the physics of the system is restricted to the lowest Bloch band and can be described by a hopping amplitude $J$ and onsite-interaction energy $U$ using the Bose-Hubbard model~\cite{fisher1989,jaksch1998}
\begin{eqnarray}\label{eq:bh-hamiltonian}
\hat H &=& -J \sum_{\langle i,j \rangle} \left(\hat a^\dagger_i \hat a_j + {\rm h.c.}\right)\nonumber\\
        && +\frac{U}{2} \sum_{i} \hat n_i(\hat n_i - 1) - \sum_{i} \mu_i  \hat n_i\,,
\end{eqnarray}
where $\hat a^\dagger_i$ ($\hat a_i$) creates (annihilates) a particle on the site with index $i=(i_x,i_y,i_z)$, $\hat n_i = \hat a^\dagger_i \hat a_i$ counts the number of particles on site $i$ and $\langle i,j \rangle$ denotes the sum over next neighbours only. The quantity $\mu_i=\mu - \epsilon_i$ is a difference between the chemical potential $\mu$ and the confining potential on corresponding site: $\epsilon_i=(m/2)\sum_{\alpha = x,y,z}  \omega_\alpha^2 d_\alpha^2 i_\alpha^2$, with $m$  the particle mass, $\omega_\alpha = 2\pi \times 20 - 60\,{\rm Hz}$ the trap frequency, and $d_\alpha=\lambda_\alpha /2$ the lattice spacing in the direction $\alpha $ defined by the laser wavelength  $\lambda_{\alpha}$. We thus work in the regime of local density approximation, when $\mu_i$ plays the role of the local chemical potential, the total number of particles $N = \sum_i n_i$ being controlled by the global chemical potential $\mu$. In the present work, the external confinement is caused by a magnetic trap and the Gaussian laser beams creating the optical lattice. 

\emph{Phase Diagram of the Trapped System}
The phase diagram for the homogeneous system (i.e. $\epsilon_{x,y,z}=0$) at unity filling is shown in Fig.~1 as a function of interaction strength $U/J$ and temperature $T$. For $T=0$, the system undergoes a quantum phase transition (QPT) from the SF to the MI phase~\cite{fisher1989, jaksch1998,sachdevbook} at the critical interaction strength $(U/J)_c = 29.34(2)$ ~\cite{capogrosso2007}. For interactions $U/J < (U/J)_c$, a phase transition between the SF and the NF exists at a critical temperature $T_c$ which tends to zero as $U/J \rightarrow (U/J)_c$. This behaviour at finite temperatures can be seen as a generic feature of QPTs, originating from a fundamental change in the ground state of the system. In this paper, we focus on the transition between SF and NF phases and the downshift of $T_c$.

\begin{figure}
\begin{center}\includegraphics[width=0.45\textwidth]{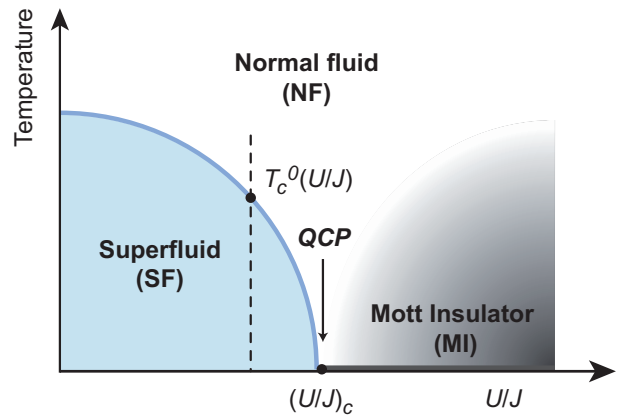}\end{center}
\caption{Simplified scheme of the finite $T$ phase diagram for a single species of bosons in a lattice potential at density $n=1$. At $T=0$, the system undergoes the transition from a SF to a MI at the critical interaction strength $(U/J)_c$. For $U/J < (U/J)_c$, the SF phase exists up to a critical temperature $T_c$ which decreases to zero at the QPC, signaling the drastic change in the ground state of the system. The MI phase right of the QCP exists strictly speaking only at $T=0$. However, Mott-like features can be observed at finite temperatures $T \ll U/k_B$. The dashed line represents a typical trace in the phase diagram along which experimental data and simulations were taken in order to determine $T_c$.}\label{fig:sfmi}
\end{figure}

In the presence of an external confinement, the system becomes inhomogeneous, rendering the notions of global 'phase diagram' or 'transition' problematic. The decrease of the local chemical potential $\mu_i$ with distance from the trap center results in the coexistence of different phases in the trap (see Fig.~2a,b), complicating the interpretation of the experimental data in general. Our strategy to avoid this ambiguity is to work with a total number of particles $N$ such that the central density stays close to unity. In order to achieve this, we determine the chemical potential $\mu \equiv \mu_1$ corresponding to a central density $n=1$ at $T = T_c(U/J)$ for a given value of $U/J$ (see Fig.~2c). A full QMC study using $\mu_1$ and including the external trap yields the target particle number $N$ which we maintain throughout the simulations and experiments. As long as $U/J<26.7$ ($T_c/J > 3$), this ensures that the SF starts to form first in the center of the trap (see Fig~2a,b). This approach justifies a direct comparison of the critical temperatures measured for the inhomogeneous system with the ones of the homogeneous system at unity filling which are well known from QMC studies~\cite{capogrosso2007}.

\begin{figure}
\begin{center}\includegraphics[width=0.45\textwidth]{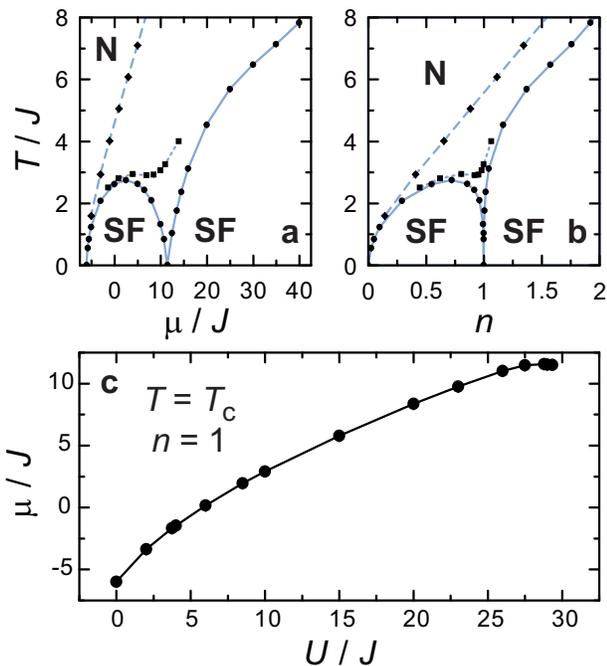}\end{center}
\caption{Phase diagram of the Bose-Hubbard Hamiltonian as a function of temperature $T/J$ and chemical potential $\mu/J$ ({\bf a}) or density $n$ ({\bf b}) at the critical interaction strength $(U/J)_c=29.34(2)$ (solid line, circles), at $U/J=26$ (dotted line, squares) and at $U/J=8.5$ (dashed line, diamonds). At the critical value $(\mu/J)_c =11.50(5)$ ($n=1$), the tip of the Mott lobe is reached where $T_c = 0$. In the lower panel ({\bf c}), we plot $\mu/J$ corresponding to a density of $n=1$ at $T=T_c$ as a function of $U/J$.}
\label{fig:sfnormal_mu}
\end{figure}

\emph{Experimental Sequence and QMC Simulations.}
Our experimental sequence starts with a Bose-Einstein condensate (BEC) of $^{87}$Rb atoms produced in a cigar-shaped magnetic trap by evaporative cooling. The condensates have no discernible thermal fraction and contain a variable number of $N = 9 \times 10^4$ to $3 \times 10^5$ atoms. The temperature is subsequently varied by exposing the BEC to a controlled and calibrated heating sequence, which allows to access initial temperatures $T_i$ between 20~nK and 400~nK while keeping $N$ constant (see Methods). After setting the temperature of the gas, we adiabatically decompress the magnetic trap towards an expanded, almost spherical trap with radial and axial trap frequencies of $2\pi \times 18.31(1)\,{\rm Hz}$ and $2\pi \times 11.69(1)\,{\rm Hz}$, respectively. Subsequently, the three-dimensional optical lattice is ramped up within $t_{\rm ramp} = 300\,{\rm ms}$ to the final depth $V_0$ using an $s$-shaped ramp~\cite{gericke2007}. The orthogonal retro-reflected laser beams forming the optical standing waves have wavelengths of $\lambda_x = 765\,{\rm nm}$ along one direction and $\lambda_{y,z} = 844\,{\rm nm}$ along the other two~\cite{latticenote}. Finally, the atoms are released by simultaneously switching off the magnetic trap and the lattices and probed after $t_{TOF}=15.5\,{\rm ms}$ using resonant absorption imaging. This procedure yields the integrated column-density $n_{\perp}(x,y)=\int n_{TOF}(\vec r)\,dz$ which is related to the single-particle density matrix in the trap.

Numerically, the Hamiltonian Eq.~(\ref{eq:bh-hamiltonian}) can be effectively simulated by the QMC worm algorithm~\cite{prokofev1998,pollet2007}. This is a statistically exact method, scaling linearly with the system volume and the inverse temperature. We can deal with realistic system sizes (up to $\sim 220^3$) at the experimentally relevant temperatures ($k_B T \leq 6J$). For a given lattice depth $V_0$, we calculate the Hamiltonian parameters $J$ and $U$ from the single particle band structure~\cite{scatnote}. The external trap parameters $\epsilon_\alpha$ are deduced from the lattice depth, the measured laser waists and magnetic trap frequencies. Since all parameter values are taken directly from the measured experimental control parameters, this comprises a full ab-initio study. The simulation results are translated into integrated column densities after TOF, taking into account the finite expansion time as well as a finite imaging resolution~\cite{gerbier2008}. The latter is accomplished by convolution of the simulation images with the Gaussian point-spread function ($5.6\,{\mu m}$ root-mean-square width) of our imaging system. This value was determined experimentally by measuring the autocorrelation function of expanding atom clouds deep in the MI regime~\cite{fölling2005}.

\emph{Comparison of QMC simulations and experimental results.}
Two typical sequences of experimental TOF images at lattice depths of $V_0 = 8\,E_r$ ($U/J=8.11$) and $V_0 = 11.75\,E_r$ ($U/J=27.5$) and for different temperatures are shown in Fig.~3 together with the corresponding simulation results. To estimate the temperature of the ensemble in the lattice potential, we start from the initial temperature $T_{i}$ measured in the magnetic trap. Using numerical data for the canonical energies and entropies in the initial and final potential, we assign an entropy $S_i(T_i)$ to a particular experimental run. The final temperature is found by inverting $T_f=T_f(S_i)$, assuming the initial entropy $S_i(T_i)$ is conserved during the lattice loading.

\begin{figure}
\begin{center}\includegraphics[width=0.45\textwidth]{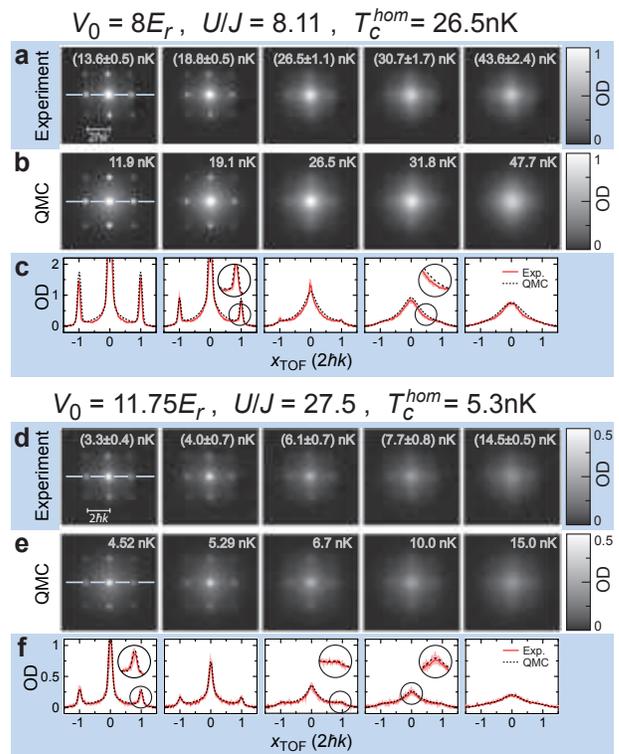}\end{center}
\caption{Comparison of experimental and simulated TOF distributions: We show the integrated column density $n_{\perp}(x,y)$ represented by the optical density (OD) as obtained from the experiment and the QMC simulations for different temperatures and two lattice depths $V_0 = 8\,E_r$ ($U/J = 8.11$, $N=2.8\times 10^5$, {\bf a-c}) and $11.75\,E_r$ ($U/J=27.5$, $N=0.9\times 10^5$, {\bf d-f}). The simulation results {\bf b, e} are selected to match the corresponding experimental distribution {\bf a, d} as close as possible and the agreement is underlined in the profiles {\bf c, f} taken along the axis denoted in the TOF distributions. The red shaded regions in {\bf c,f} represent the experimental peak-to peak fluctuations of the TOF densities, while the red solid line is an average over at least three experimental shots. To the experimental images, we assign the temperatures $T_f$ calculated for an adiabatic loading process (see text) where the errors stem from the uncertainty in the calibration of the initial temperature $T_i$. The temperatures $T_f^\prime$ for the QMC results are exact. The comparison at $V_0=8\,E_r$ confirms the adiabaticity of the loading process, while for $V_0=11.75$ we find a small general shift in temperature by up to 30\% due to heating processes.}\label{fig:TOFseries}
\end{figure}

Besides relying on the ``adiabatic'' temperature $T_f$, another possibility is to match the experimental profile with that computed from QMC simulations. This yields a ``matching'' temperature $T_f^\prime$. Since the temperature used for the simulations is exact, this procedure can be seen as a direct thermometry for the experiment. In our comparison, the accuracy is limited by the sampling of the experimental and simulation data along the temperature axis. For $V_0 = 8\,E_r$, we find very good agreement between the measured and simulated TOF distributions for $T_f = T_f^\prime$. However, for $V_0 = 11.75\,E_r$ we find the best agreement for $T_f^\prime$ being slightly higher (up to a maximum of 30\%) than $T_f(S_i)$. We interpret this shift in the final temperature as a signal for non-adiabatic heating of the ensemble during the loading of the lattice due to spontaneous scattering of lattice photons, technical noise in the laser setup or the finite ramping times. We can estimate the expected heating effect due to the spontaneous scattering of lattice photons and fluctuations of the dipole force (see Methods). For the two final lattice depths of $8\,E_r$ and $11.75\,E_r$ presented in Fig.~3, we find a temperature increase at $T=T_c$ by 3.1(3)\% and 30(3)\%, respectively, which can already quantitatively explain the observed shift in the latter case. We note, that at the lowest initial temperatures the quantum gas remains well in the degenerate regime throughout the parameter range investigated.

For the smallest values of $T_f$ and $T_f^\prime$, we generally observe sharp interference peaks separated by $2\hbar k$ on top of a broader pedestal pattern (Fig.~3). The weight of these peaks decreases with temperature until only the pedestal is left, which blurs further as the temperature is increased even more. In three dimensions, the existence of a superfluid requires the existence of a BEC, i.e. long-range order~\cite{leggett1999}. This long-range phase coherence gives rise to the narrow peaks in the TOF distributions. We therefore use the sudden onset of sharp interference peaks (or a sudden change in their width) to identify the onset of condensation and superfluidity in the lattice~\cite{kollath2004}. The width $w_p$ of the sharp peaks is limited in principle by the coherence length $l_c$ of the BEC in the trap according to $w_{p}\approx \hbar t/m l_{c}$. In practice, the finite expansion time $t_{TOF}$ and imaging resolution do not allow to measure $l_{c}$ accurately when long-range order is present. In agreement with refs.~\cite{kashurnikov2002, kato2008, zhou2009}, we find that interference peaks can also be observed in the background pattern. However, due to the short coherence length in the normal phase, these peaks are typically much broader than the ones caused by the long-range phase coherent superfluid observed at lower values of $U/J$, and vary smoothly in temperature throughout the parameter range investigated here. 

The following evaluation steps have been carried out simultaneously for the experimental and the QMC data. In general, this evaluation does not require the simulation results which we use to verify the method. In order to find a suitable indicator for the boundary between the SF and NF phases, we fit both the experimental and the QMC data to an empirical model function
\begin{eqnarray}\label{eq:model}
  n_\perp(x,y) = W(x,y) \left[n_{c}(x,y) + n_{bg} + n_{th}(x,y)\right]\,,
\end{eqnarray}
where the envelope $W(x,y)$ represents the momentum-space Wannier function, approximated by a Gaussian. Here, we assume that the atomic distribution can be split into three contributions, a ``condensed component'' $n_{c}(x,y)$ modeled by Gaussian peaks of width $w_p$ at the position of the reciprocal lattice nodes, a constant background $n_{bg}$ modeling an ``incoherent component'' 
and a ``thermal component'' $n_{th}(x,y)$ modeling thermal excitations of the condensate. The latter term is chosen as $n_{th}(x,y)  = \sum_{l=1}^{l_{max}} (\exp[\cos(k_xx)+\cos(k_yy)-2])^{J \beta l}/l^{3/2}$. In the limit $l_{max} \to \infty$, this corresponds to the analytical solution for noninteracting bosons in a lattice potential with temperature $T = 1/k_B\beta$. We truncate the sum at $l_{max}=4$ as a compromise to achieve small residuals and keeping the numerical effort small at the same time. Here, we focus on the fraction of atoms in the Gaussian peaks $f_p$ and their width $w_p$ alone~\cite{parameternote}.

The fit results for three sets of simulations and experiments are displayed in Fig.~4 together with the visibility of the interference patterns, computed as in ref.~\cite{gerbier2005a}. Although the fit function Eq.~\ref{eq:model} is able to reproduce the TOF profiles quite well, there is some ambiguity in the interpretation of the various components given above. For $U/J < 20$, the thermal and condensed parts are well accounted for by the terms $n_{th}(x,y)$ and  $n_{c}(x,y)$. Because of our finite momentum resolution, the width $w_p$ of  $n_{c}(x,y)$ is practically constant with temperature up to $T=T_c^{\rm peak}$, and the condensed fraction is well evaluated from the peak amplitude $f_{p}$. We find a sudden onset of the fraction of atoms in peaks $f_p$ as the temperature is lowered, where $f_p$ increases approximately linearly from this point. We fit two lines to the data where one is the horizontal axis at $f_p=0$. The intersection point gives an estimate $T_c^{\rm peak}$ of the critical temperature, as was done for a harmonically trapped Bose gas in ref.~\cite{gerbier2004a}. At values $U/J > 20$, the thermal component is not fully captured by $n_{th}$ due to interactions between non-condensed atoms. To compensate for this, the fit finds non-zero amplitude for $n_c$ even in the absence of sharp peaks above $T_c$. However, we find $w_p$ to stay constant until a certain temperature $T < T_c^{\rm peak}$ above which it suddenly increases, corresponding to a decreasing coherence length. We assign this increased peak width (which can be seen directly in Fig.~3) to the thermal component of the lattice gas. Again, we fit two lines to the data and extract a second intersection point at $T_c^{\rm width} < T_c^{\rm peak}$.

\begin{figure}
\begin{center}\includegraphics[width=0.45\textwidth]{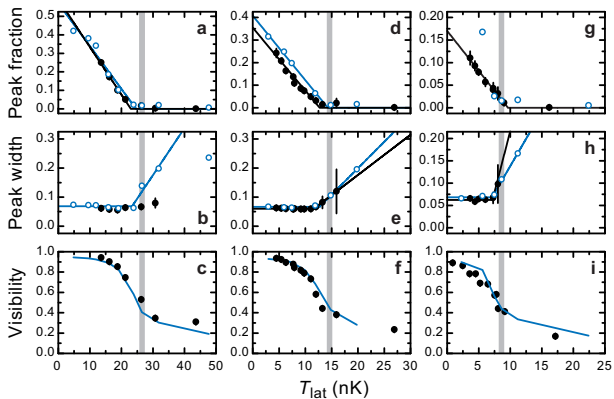}\end{center}
\caption{Fit results for the onset of superfluidity: We plot the peak fraction $f_p$ (top row), the peak width $w_p$ (center row) and the visibility (bottom row) for lattice depths of $8 E_r$ ({\bf a-c}, $U/J = 8.11$), $10\,E_r$ ({\bf d-f}, $U/J=16.1$) and $11.25\,E_r$ ({\bf g-i}, $U/J=23.7$). Both, the results for the experimental data (filled circles) and the QMC simulations (open circles) are presented. The solid lines in the upper two rows correspond to the fits made in order to obtain an estimate for $T_c$. The grey vertical lines indicate $T_c$ for a homogeneous lattice at unity filling~\cite{capogrosso2007}.}\label{fig:fitresults}
\end{figure}

We plot $T_c^{\rm peak}$ and $T_c^{\rm width}$ as obtained from the QMC simulations and the experimental data as a function of $U/J$ in Fig.~5. Here, the experimental data points are shifted in temperature to account for the heating during the lattice loading (see Methods). Both experimental and QMC results consistently exhibit a suppression of $T_c$ when the interaction strength is increased towards its critical value, as it has been predicted for the homogeneous Bose-Hubbard model~\cite{dickerscheid2003a,capogrosso2007}. Above $U/J=20$, as discussed above, this behaviour cannot be extracted from the fraction of atoms in sharp peaks due to the inability of the fit function to separate the thermal and condensate contributions in this parameter range. Instead, it must be inferred from the analysis of the peak width, reflecting the spatial coherence properties of the lattice gas. On the other hand, the width alone does not yield enough information for lower lattice depth, since the fit routine assigns zero amplitude above $T_{c}^{\rm peak}$. We thus have to use a combination of both methods in order to determine $T_c$, which are found to overlap for intermediate values of $U/J$ (see Fig.~5). The critical temperatures obtained from the inhomogeneous system in the experiments and simulations are remarkably close to the ones of the homogeneous case with unity filling (solid line in Fig.~5~\cite{capogrosso2007}), apart from a small ($\simeq 10-15\%$) systematic shift towards lower values. This shift is almost entirely due to applying our evaluation procedure to the inhomogeneous density cloud in the strongly interacting limit.

\begin{figure}
\begin{center}\includegraphics[width=0.45\textwidth]{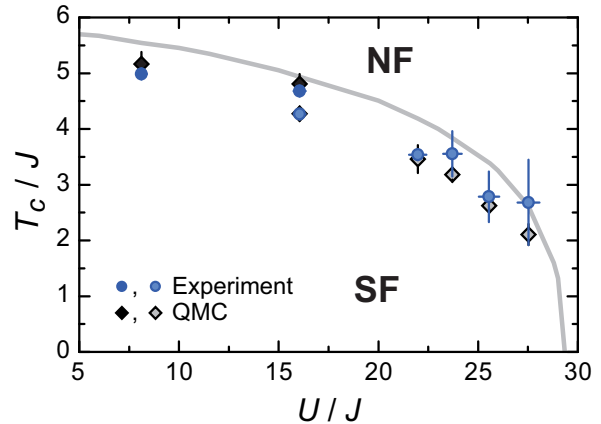}\end{center}
\caption{Finite temperature phase diagram and suppression of $T_c$ in the lattice. The critical temperature for superfluidity $T_c/J$ as obtained from the measurement (circles) and the QMC simulations (diamonds) is plotted versus the interaction strength $U/J$. Closed symbols mark the values $T_c^{\rm peak}$ extracted from the peak fraction $f_p$, while the values represented by open symbols represent $T_c^{\rm width}$ as obtained from the evaluation of the peak width $w_p$. The experimental data has been corrected to account for heating during the lattice loading (see text). The solid line is the QMC result for the homogeneous Bose-Hubbard model at unity filling, taken from ref.~\cite{capogrosso2007}.}\label{fig:Tc}
\end{figure}

\emph{Conclusions and Outlook}
In conclusion, we present for the first time a full quantitative comparison between experiment and ab-initio quantum Monte Carlo simulations for large ($N \simeq 10^5$) systems of ultracold bosons in optical lattices. Using only experimentally measured parameters as input to the simulations and assuming adiabatic loading into the lattice, we find remarkable agreement up to $U/J \simeq 20$. Discrepancies in the final temperature of up to 30\% are observed for deeper lattices, which are resolved by accounting for specific heating mechanisms. The direct comparison of experimental and simulated TOF images allows us to perform accurate thermometry for interacting bosons in an optical lattice. Up to $U/J \simeq 20$, we find that for typical parameters the sudden appearance of sharp interference peaks with increasing temperature yields a reliable measure for the onset of superfluidity. For larger interaction strengths, very close to the quantum critical region, we observe a smoother transition to the NF phase, including broad interference features for thermal samples. In this parameter range, we find that the change in the width of the interference peaks gives more reliable information for the determination of the critical temperature $T_c$. Using these analysis techniques, we observe the suppression of $T_c$ upon approaching the QCP for the SF-MI transition in both the experimental and the simulated data, thus mapping out the finite temperature phase diagram of the system. Furthermore, we find that up to $U/J\simeq 27.5$ the bosonic gas remains well in the degenerate regime for the lowest initial entropies used, in contrast to the theoretical analysis presented in ref.~\cite{diener2007}. Our results demonstrate the potential of using ultracold atoms in optical lattices to quantitatively study large-scale condensed-matter physics. The direct measurement of the suppression of $T_c$ may furthermore open the way to approach the region above the QCP to experimentally study quantum critical phenomena.

We would like to thank B. Capogrosso-Sansone, P. N. Ma, F. C. Zhang, S. F\"olling, H. Moritz, T. Esslinger and J. Dalibard for stimulating discussions. This work was supported by the DFG, the SNF, the EU (IP SCALA), DARPA (OLE program) and AFOSR. The simulations were run on the Brutus cluster at the ETH in Z\"urich.

\section*{Methods}
\subsection{Controlled heating sequence.}
One important requirement for the presented measurements is the ability to change the initial temperature $T_i$ of the atomic ensemble without changing the number of particles $N$. We use a one-dimensional optical lattice with wavelength $\lambda_y = 844\,{\rm nm}$ superimposed to the magnetic QUIC trap and perpendicular to its slow axis to transfer energy to the ensemble in a controlled way. After evaporative cooling, we slowly ramp this lattice to a final value $V_{\rm heat}$ using a $s$-shaped ramp~\cite{gericke2007}, before we rapidly pulse it off and on for four times using linear ramps of $1 \,{\rm ms}$ length. With the lattice at its high value, the created excitations are let to thermalize with the rest of the sample over a holdtime of $800\,{\rm ms}$, before the lattice is ramped down again within $700\,{\rm ms}$. After this heating sequence, we find no significant reduction of the particle number and no residual sloshing could be observed.

\subsection{Temperature measurement in the magnetic trap.}
To measure the temperature $T_i$ of the ensemble after the heating sequence, we release the cloud from the QUIC-trap and probe its TOF distribution after $18.5\,{\rm ms}$ of free expansion by resonant absorption imaging. The integrated column density $n_\perp(x,y)$ is fitted by a bimodal distribution~\cite{ketterle1999a}
\begin{eqnarray}
  \tilde n(\rho,z) &=& \frac{6 n_{th}^0}{\pi^2}\,g_2\left[\exp(1-\rho^2/\sigma_\rho^2 - z^2/\sigma_z^2)\right]\nonumber\\
  +\tilde n_{BEC}(\rho,z)
\end{eqnarray}
where $\rho$ and $z$ are the radial and axial coordinates with respect to the orientation of the QUIC-trap in the imaging plane, $g_{\alpha}(x)\equiv \sum_n z^n/n^\alpha$ is the Bose function and $\tilde n_{BEC}(\rho,z)$ is the Thomas-Fermi distribution of the condensed part. From the radial and axial radii $\sigma_\rho$ and $\sigma_z$ of the thermal component and the trap frequencies $\omega_\rho = 2\pi \times 130.4(2)\,{\rm Hz}$ and $\omega_z = 2\pi \times 16.3(1)\,{\rm Hz}$, we calculate the radial and axial temperatures of our sample as~\cite{ketterle1999a}
\begin{equation}
  k_{\rm B} T_i^{\rho,z} = \frac{1}{2}\,m\left(\frac{\omega_{\rho,z}^2}{1+\omega_{\rho,z}^2 t^2}\,\sigma_{\rho,z}^2\right)\,,
\end{equation}
Both temperatures differ by maximally a few percent, indicating thermalization. To calibrate our heating sequence, we measure $T_i$ as a function of $V_{heat}$ for each total number of particles $N$ used in the experiments.

\subsection{Adaptation of the tunnel coupling for the bichromatic lattice.}
In the experiments we ensure that the redistribution of the atoms happens with equal tunnel coupling in all three spatial directions of our bichromatic lattice. Based on the numerically calculated tunnel couplings, we choose the lattice depth in the $x$-direction to be $V_x=1.24\,V_{y,z}+0.80\,E_r$ throughout the loading of the lattice, where $V_{y,z}$ are the lattice depths of the $y$- and $z$-directions.

\subsection{Estimation of the heating power in the optical lattice.}
We estimate the heating rate in the optical lattice potential following the argument of momentum-diffusion in ref.~\cite{gordon1980} for an atom at rest in a standing wave light field. Taking into account the contributions from spontaneous emission and the fluctuating dipole force, one finds a position independent heating rate $\langle\dot E\rangle \simeq E_r \Gamma s_{\rm max} /2$, where $\Gamma$ is the lifetime of the excited state of the atom, $s_{\rm max} \approx 2 \Omega^2/\Delta^2$ is the saturation parameter at an intensity maximum, $\Delta$ is the detuning of the laser light and $\Omega$ is the coupling Rabi frequency. Given the potential depth $V_{\rm SW}=\hbar \Delta s_{\rm max} /2$, this translates into $\langle\dot E\rangle \simeq E_r \Gamma_{\rm sc}$ with the scattering rate $\Gamma_{\rm sc} = V_{\rm SW} \Gamma /\hbar \Delta$.

By comparing QMC simulations to experimental TOF images taken after loading of the lattice and a subsequent variable holdtime $t$ at a fixed lattice depth $V_{0}$, we directly measure the increase in the energy per particle versus $t$. From repeated measurements at $V_0 = 8,10$ an $12\,E_r$, we obtain $\langle\dot E\rangle = 0.59(8) ~E_r \Gamma_{sc}$. When calculating $\Gamma_{\rm sc}$, we take into account the different detunings $\Delta_{x,y,z}$ of our lattice beams and the fine splitting of the atomic resonance into $D_{1}$ and $D_{2}$ lines. The reduced measured heating power could hint that the excitations created due to the heating do not fully thermalize with the rest of the ensemble on the timescale of the experiment, possibly due to reduced collisions between atoms populating different Bloch bands~\cite{challis2009a}. A quantitative, first principle calculation of the heating rate is beyond the scope of this paper, where we use the measured value.

For the measurements presented in the main text, we estimate the total heating by assuming the relation $\langle\dot E\rangle = 0.59~ E_r \Gamma_{sc}$ to be valid for any lattice depth and making use of the almost point symmetric shape of the ramp profile~\cite{gericke2007}. We find a total increase in energy by $9(1)\times10^{-3}\,E_r$ for $V_0 = 8\,E_r$ and $12(2)\times 10^{-3}\,E_r$ for $V_0 = 11.75\,E_r$. Using numerical data for the canonical energy in the final lattice potential, we obtain an increase in temperature near $T=T_c$ by 3.1(3)\% and 30(3)\% in the two respective cases.


\bibliography{FinTBib}

\begin{thebibliography}{10}
\bibitem{latticenote}
 \bibinfo{note}{The laser intensities for the two different wavelengths have
  been adapted to obtain the same value of $J$ along all three directions. The
  lattice depths are given in units of recoil energy $E_R =
  h^2/(2m\lambda_{y,z}^2)$ for the longer period lattices.}
\bibitem{scatnote}
 \bibinfo{note}{To calculate $U$, we use a $s$-wave scattering length of $a_s =
  5.31\,{\rm nm}$ for atoms in the $|F=1,m_F=-1\rangle$ Zeeman state. The
  Gaussian shape of the lattice beams in principle leads to a decrease of $V_0$
  and thus $U/J$ from the trap center towards the outer regions of the sample.
  Since the effect on the TOF distribution is on the percent level only, we
  work with constant values for $U$ and $J$.}
\bibitem{parameternote}
 \bibinfo{note}{Although this model function reproduces well both the
  experiments and the simulations, the fitted parameter $\beta$ was found to
  deviate significantly from the temperature assigned through the
  entropy-matching procedure, making it unreliable for thermometry. This can be
  expected for strong interactions since the term $n_{th}(x,y)$ in
  Eq.~(\ref{eq:model}) is derived from the solution for non-interacting
  Bosons.}
\end{thebibliography}


The authors declare that they have no
competing financial interests.

Correspondence and requests for materials
should be addressed to S. Trotzky~(email: trotzky@uni-mainz.de).


\end{document}